\documentclass[twocolumn,aps,prl]{revtex4-1}

\usepackage{slashed}
\usepackage{tikz-cd}
	\usepackage{amsmath}
	\usepackage{amsfonts}
	\usepackage{amssymb}
	\usepackage{graphicx}
	\usepackage{mathtools}
	\usepackage{stmaryrd}
	\usepackage{bbm}
	\usepackage[colorlinks=true,citecolor=blue,linkcolor=red]{hyperref}
	\usepackage{bbold}					
	\usepackage[makeroom]{cancel}		
	\usepackage{multirow}				
	\usepackage[normalem]{ulem}        
	\usepackage{array}
	
\usepackage{framed}

	\newcolumntype{x}[1]{>{\centering\let\newline\\\arraybackslash\hspace{0pt}}p{#1}}
	
	\makeatletter
\newcommand*\rel@kern[1]{\kern#1\dimexpr\macc@kerna}
\newcommand*\widebar[1]{%
  \begingroup
  \def\mathaccent##1##2{%
    \rel@kern{0.8}%
    \overline{\rel@kern{-0.8}\macc@nucleus\rel@kern{0.2}}%
    \rel@kern{-0.2}%
  }%
  \macc@depth\@ne
  \let\math@bgroup\@empty \let\math@egroup\macc@set@skewchar
  \mathsurround\z@ \frozen@everymath{\mathgroup\macc@group\relax}%
  \macc@set@skewchar\relax
  \let\mathaccentV\macc@nested@a
  \macc@nested@a\relax111{#1}%
  \endgroup
}
\makeatother

	\DeclareMathAlphabet{\mathbbold}{U}{bbold}{m}{n}

	\def\bra#1{\left<{#1}\right|}				
	\def\ket#1{\left|{#1}\right>}				

	\def\PRLgreater{\,{>}\,}
	\def\PRLless{\,{<}\,}
	\def\PRLequal{\,{=}\,}
	\def\PRLminus{\,{-}\,}

	\newcounter{subeqn} %
	\makeatletter
	\@addtoreset{subeqn}{equation}
	\makeatother

\definecolor{TB}{rgb}{0,0,0} 
\definecolor{XQ}{rgb}{1,0,0}
\definecolor{CW}{rgb}{.5,0,.5}

\newcommand{\beginsupplement}{%
        \setcounter{table}{0}
        \renewcommand{\thetable}{S\arabic{table}}%
        \setcounter{figure}{0}
        \renewcommand{\thefigure}{S\arabic{figure}}
        \setcounter{equation}{0}
        \renewcommand{\theequation}{S\arabic{equation}}%
     }
\newcommand{\inp}[2]{\langle#1|#2\rangle}

\begin{document}
\title{Multi-gap topology of the Wilson loop operator in mirror symmetric insulators}
\author{Penghao Zhu}
\author{Taylor L. Hughes}
\author{Xiao-Qi Sun}\email{xiaoqi20@illinois.edu}
\affiliation{Department of Physics and Institute for Condensed Matter Theory,
University of Illinois at Urbana-Champaign, Illinois 61801, USA}
\date{\today}

\begin{abstract}
We study the multi-gap topology of the periodic spectra of Wilson loop operators (WLOs) in mirror symmetric insulators. We develop two topological invariants each associated with a mirror-invariant gap in the Wilson loop spectrum. We propose that both topological invariants in combination determine the general higher-order bulk-boundary correspondence in 2D mirror symmetric, boundary-obstructed topological insulators. Finally, we demonstrate that these new multi-gap topological invariants apply to anomalous cases beyond those captured by the nested Wilson loop, and we subsequently develop an understanding of the correlation between WLOs along two orthogonal directions.
\end{abstract}

\maketitle

 Topological band theory has been instrumental in describing and predicting symmetry-protected topological phases of electrons \cite{qi2011topological,bansil2016colloquium,bradlyn2017topological,cano2018building}. The predominant idea behind the theory is to study the topology of Bloch Hamiltonians in the presence of an energy gap. It was not until the recent introduction of higher-order topological phases \cite{benalcazar2017electric,benalcazar2017quantized,khalaf2021boundary,schindler2018higher,song2017d,josias2017reflection,schindler2018higher2,serra2018observation,peterson2018quantized,noh2018topological,imhof2018topolectrical,ni2019observation,xue2019acoustic,benalcazar2019quantization,koji2019higher,zhang2019higher,yan2019higher,you2018higher,dubinkin2019higher,park2019higher,zhang2020higher,sayed2020higher,li2020fractional,peterson2020fractional,zhu2021quantized,aggarwal2021evidence} that the topology of \emph{gapped} Wilson loop operators (WLOs) attracted attention \cite{benalcazar2017quantized,benalcazar2017electric,khalaf2021boundary,schindler2018higher}. 
Previous work has argued that among the possible gaps of the WLO spectrum, there is a 
\emph{specific} gap that directly 
corresponds to an energy gap of the boundary of a system. Furthermore, it has been proposed that topological invariants defined for this gap characterize a fundamentally new type of topological phase having a topological obstruction protected by the boundary gap \cite{benalcazar2017quantized,benalcazar2017electric,khalaf2021boundary}. 

\begin{figure}[h]
    \centering
    \includegraphics[width=1\columnwidth]{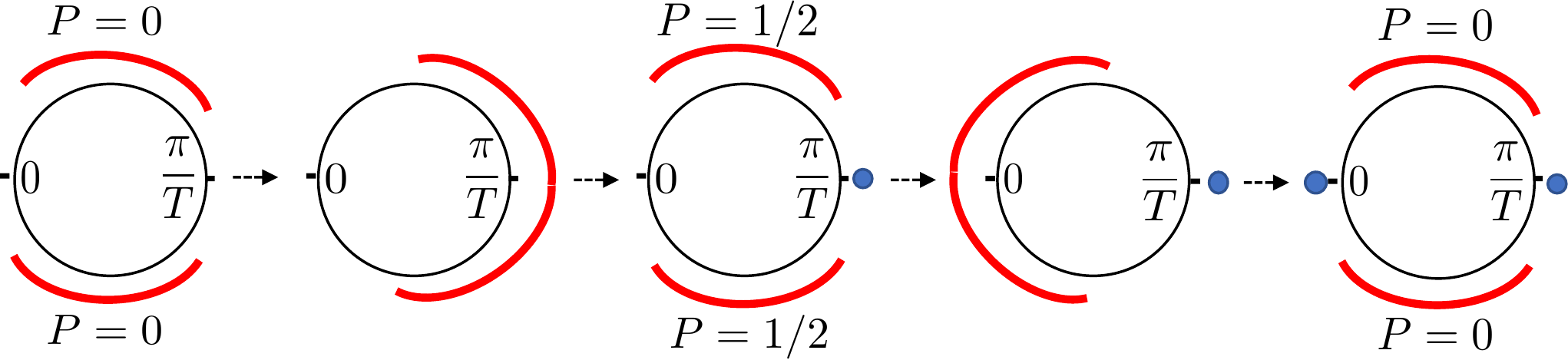}
    \caption{
    As an analogy between Floquet systems and the topology of the Wilson Loop Operator we show a series of gap closing processes of Floquet bands (red arcs) in the quasienergy space (black circles). From left to right we show that a gap closing at $0$ and a gap closing at $\pi/T$ (for Floquet period $T$) generates in-gap states (indicated by blue dots) at $0$ and $\pi/T$, while the polarization (Zak-Berry phase over the Brillouin zone) changes from $0$ to $1/2$ and then back to $0$. Thus, in the final case, the correspondence between $P=0$ and the existence of in-gap boundary modes is broken.  }
    \label{fig:floquetanomaly}
\end{figure}

This approach to topological classification treats the eigenstates of the WLO analogously to energy states of an effective band insulator. However, since the spectrum of the (unitary) WLO is \emph{periodic}, it naturally suggests a \emph{multi-gap} description of the topological invariants. A similar situation occurs in cyclically driven Floquet systems that exhibit a periodic quasi-energy spectrum. A multi-gap classification of the topology of Floquet bands predicts novel anomalous topological states whereas static analogs fail to describe both the bulk topology and the boundary correspondence (see an illustration in Fig.~\ref{fig:floquetanomaly})\cite{rudner2013anomalous,asboth2014chiral,fruchart2016complex,zhang2020theory}. The clear analogy between the periodic spectra of a WLO and a Floquet operator necessitates a multi-gap topological classification of the former.  Indeed,  examples of anomalous higher order topological insulators have already been presented~\cite{Franca:2018} which could be described by a framework that incorporates the multi-gap classification scheme. 

To this end, here we explore the multi-gap topology of WLOs for 2D systems with mirror symmetries along both the $x$ and $y$-directions. Because of the connection between the WLO eigenvalues and the positions of Wannier centers \cite{kivelson1982wannier,marzari1997maximally,resta1998quantum}, for directions with mirror symmetry there are two distinguished WLO eigenvalues (corresponding to two distinguished Wannier positions) which are mirror-invariant, i.e., the values $0$ and $1/2$. For our cases of interest, the WLO spectrum is gapped at both $0$ and $1/2,$ and we can define two sets of topological invariants associated with 
the gaps at $0$ and $1/2$, respectively. We show that  both sets of topological invariants, in combination, determine the number, and location, of  Wannier orbitals at the boundary of the system, and thus determine the boundary topological obstruction. For 2D mirror symmetric insulators our results hence describe a general multi-gap topological classification of the WLO and its boundary correspondence, i.e., a higher-order bulk-boundary correspondence (HOBBC).

Our multi-gap topological formalism also clarifies two important aspects of the HOBBC. First, our method provides a general prescription to describe the number of in-gap modes of open boundary WLOs at $0$ and $1/2$, including the previously reported anomalous cases, that were not captured by the simple nested Wilson loop analysis. Indeed, we establish  two $\mathbb{Z}$ topological invariants to capture in-gap modes of WLOs in both distinguished gaps. Crucially our invariants recover the additive group structure expected for topological invariants when adding extra occupied bands, which is absent in the nested Wilson loop formalism. 
Furthermore, our formalism lays plain the correlation of WLOs along two orthogonal directions, e.g., in the BBH model the nested Wilson loop along one direction is related to the in-gap boundary modes of the WLO along another orthogonal direction.  
Thus, our framework offers a new route for the study of the topology of WLOs and the HOBBC. 

\emph{WLO and its spectrum.--} Before proceeding to the discussion of the multi-gap topology of a WLO, we first review the definition of a WLO and its spectrum for a band insulator. Let $P(\mathbf{k})$ be the projector onto the occupied bands at $\mathbf{k}.$ We can define the Wilson \emph{line} operator as
\begin{equation}
\label{eq: WLO}
\widehat{\mathcal{W}}_{\mathbf{k}_{2}\leftarrow\mathbf{k}_{1}}=\prod_{\mathbf{k}=\mathbf{k}_{1}}^{\mathbf{k}_{2}}P(\mathbf{k}),
\end{equation}
where $\prod_{\mathbf{k}\PRLequal\mathbf{k}_{1}}^{\mathbf{k}_{2}}$ is a path-ordered product along the path indicated by $\mathbf{k}_{2}\leftarrow\mathbf{k}_{1}$. When $\mathbf{k}_{2}\PRLequal\mathbf{k}_{1}$, the Wilson line operator in Eq.~\eqref{eq: WLO} becomes a WLO. The exponents of all unit modulus complex eigenvalues of the WLO (divided by $2\pi$) are defined to be the spectrum of WLO (also called the Wannier spectrum), which form spectral bands called Wannier bands. The spectrum of the WLO gives the Wannier centers of the band electrons, and is well defined modulo the lattice constant (set to unity for simplicity) due to the translation symmetry of the lattice; and we henceforth always choose the branch cut $[0,1)$.  Apart from these Wannier bands, there are also flat, zero-modulus eigenvalue bands (ZEBs) of the WLO which are formed from the unoccupied energy bands, and they will be particularly important in our later discussion of topological phase transitions.

\emph{Boundary in-gap states of a WLO.--} The periodic and open boundary spectra of WLOs provide key clues to achieve our goal, which is the HOBBC in 2D mirror symmetric insulators. Just as a bulk gapped Hamiltonian can manifest in-gap, boundary-localized modes with open boundary conditions, so can a gapped WLO. To understand the nature of the bulk and  boundary spectra of a WLO let us consider (without loss of generality) the mirror symmetry $m_{x}$ along the $x$-direction, satisfying $m_{x}^2\PRLequal 1$ and $m_{x} P(k_{x},k_{y})m_{x}\PRLequal P(-k_{x},k_{y})$. Then, at a fixed $k_{y}$, let us consider two WLOs along the $k_{x}$-direction, each starting from a different $m_x$-invariant momentum $(0,k_{y})$ or $(-\pi,k_{y}).$ These WLOs are denoted as $\widehat{\mathcal{W}}_{x,2\pi \leftarrow 0}(k_{y})$ and $\widehat{\mathcal{W}}_{x,\pi \leftarrow -\pi}(k_{y})$, 
and they obey the symmetry relations:
\begin{equation}
\label{eq: mirrorWLO}
\begin{aligned}
m_{x}\widehat{\mathcal{W}}_{x,2\pi \leftarrow 0}(k_{y})m_{x}&= \widehat{\mathcal{W}}^{\dag}_{x,2\pi\leftarrow 0 }(k_{y}),
\\
m_{x}\widehat{\mathcal{W}}_{x,\pi \leftarrow -\pi}(k_{y})m_{x}&= \widehat{\mathcal{W}}^{\dag}_{x,\pi\leftarrow -\pi }(k_{y}).
\end{aligned}
\end{equation}
These relations enforce the spectra of both WLOs to be symmetric with respect to $0$ and $1/2$ such that eigenstates come in pairs having eigenvalues $e^{\pm 2\pi i w_{x}}$ when $w_{x}\neq 0,1/2$. As such, isolated eigenstates are locked by the mirror symmetry to sit at  either $0$ or $1/2$. 

In the following, we focus on the WLOs that have bulk gaps at $0$ and $1/2$, but which manifest protected boundary states at $0$ and/or $1/2$ for open boundaries (in the $y$-direction for $\widehat{\mathcal{W}}_{x}$). To distinguish the WLOs under different boundary conditions, we drop off the ``$(k_{y})$" when we represent the WLOs in a cylinder geometry where the $x$-direction is periodic and the $y$-direction is open, 
i.e., we use $\widehat{\mathcal{W}}_{x,2\pi \leftarrow 0}(k_{y})$ for periodic and  $\widehat{\mathcal{W}}_{x,2\pi\leftarrow 0}$ for open. Since the spectrum of the WLO identifies the locations the electronic Wannier centers (i.e., the center of localized electronic charge), the number of states at $0$ and $1/2$ tells us the number of Wannier orbitals centered the two mirror invariant $x$ positions on the sharply terminated $y$ boundary, i.e., the Wannier center configuration of the boundary. With $m_{x}$ symmetry, the boundary Wannier orbitals can either sit on the maximal Wyckoff positions at $0$ or $1/2,$ or form a non-maximal general Wyckoff position where a pair of orbitals sit at $(-x_0,y), (x_0,y)$ in a unit cell.  We call the Wannier orbitals localized at the maximal positions lying in between two boundary unit cells obstructed Wannier orbitals, and we call the Wannier orbitals localized at the center of a boundary unit cell, and the general Wyckoff pair, trivial Wannier orbitals.  If we have an odd number of obstructed Wannier orbitals associated to each unit cell, then the boundary has a quantized polarization $1/2$, and thus we say there is a boundary obstruction protected by $m_x$\cite{benalcazar2017electric,khalaf2021boundary}. Similar conclusions can be drawn for WLOs along the $y$-direction since we also have mirror symmetry $m_{y}.$

\emph{Topological invariants and HOBBC.--} We can now develop two topological invariants that capture the number of robust in-gap boundary states of the WLO for the gaps at $0$ and $1/2$ respectively. 
We can construct the two invariants from the two WLOs in Eq.~\eqref{eq: mirrorWLO} because they have  different sensitivities to the two gaps. We can show this by an analogy to chiral-symmetric Floquet systems\cite{asboth2014chiral}, because we can treat $m_{x}$ as an effective chiral symmetry in the WLO spectrum. Hence, we can define two effective chiral winding numbers: $\nu_{1}$ for $\widehat{\mathcal{W}}_{x,2\pi\leftarrow 0}(k_{y})$, and $\nu_{2}$ for $\widehat{\mathcal{W}}_{x,\pi\leftarrow -\pi}(k_{y})$ as \cite{mon2014topological}
\begin{equation}
\label{eq: chiralwinding}
\nu_{1,2}=\frac{i}{\pi} \int_{0}^{2 \pi} d k_{y}
\sum_{n\in v}\left\langle m_{x} w^{1,2}_{n}(k_{y})\left|\partial_{k_{y}}\right| w^{1,2}_{n}(k_{y})\right\rangle,
\end{equation}
where $\ket{w_{n}^{1}(k_{y})}$ ($\ket{w_{n}^{2}(k_{y})}$) is the eigenstate of $\widehat{\mathcal{W}}_{x,2\pi\leftarrow 0}$ ($\widehat{\mathcal{W}}_{x,\pi\leftarrow -\pi}$), $n$ is the band index  and $v$ represents one of the two $m_x$-related (effectively chiral related) subspaces of the WLO eigenstates (which includes Wannier bands and ZEBs). Hence, to calculate $\nu_{1,2}$ we need to separate all WLO bands into two $m_x$-related (effectively chiral related) sets, and choose one of them to be $v$ \footnote{No matter which one we choose, Eq.~\eqref{eq: chiralwinding} gives the same result.}. For each WLO $\in \{\widehat{\mathcal{W}}_{x,2\pi\leftarrow 0},\widehat{\mathcal{W}}_{x,\pi\leftarrow -\pi}\}$, the gaps in the Wilson loop spectrum naturally give a separation of 
Wannier bands into effective chiral subspaces, but we also must separate 
the ZEBs into $m_x$-related sets as well.
 To do this it is convenient to find a basis for ZEBs with definite $m_x$ eigenvalues and also with a smooth gauge~\cite{Soluyanov2012smooth} in $k_y$. We label this basis as $\{\ket{\psi^{+}_{j}(k_{y})}\}$ and $\{\ket{\psi^{-}_{l}(k_{y})}\},$ where $j\PRLequal 1,2,...,N_{+}$ and $l\PRLequal 1,2,...,N_{-}$, run over states with $m_{x}$ eigenvalue $+1$ and $-1$ respectively. We focus on the case where $N_{+}\PRLequal N_{-}\PRLequal N$, and then the two sets of mirror-related bands are formed by linear combinations of pairs such as $(\ket{\psi^{+}_{j}(k_{y})}\pm \ket{\psi^{-}_{l}(k_{y})})/\sqrt{2}.$ For each pair we choose one of the $\pm$ linear combinations to reside in $v,$ such that $v$ and its complement are related by $m_{x}$. Moreover, it can be proven (see Supplementary Material\cite{Supp}) that  the chiral winding number contribution from ZEBs always takes the form:
\begin{equation}
\frac{i}{2\pi}\int_{\text{BZ}} dk_{y}\sum_{j=1}^{N}\left(\inp{\psi^{+}_{j}}{\partial_{k_y}\psi^{+}_{j}}-
\inp{\psi^{-}_{j}}{\partial_{k_y}\psi^{-}_{j}}\right),
\label{eq: zc}
\end{equation} independent of the choice of separation. However, we note that the gauge choice of the separated bands can alter the results of the chiral winding number by an integer (e.g., a gauge transformation of $|\psi^{+}_{j}\rangle \rightarrow e^{i k_y}|\psi^{+}_{j}\rangle$). Thus, to make Eq.~\eqref{eq: chiralwinding} a well-defined topological number, we need to fix a gauge for both $|\psi^{+}_{j}\rangle$ and $|\psi^{-}_{j}\rangle$. 
In order to compare the results in different phases we require the \emph{same} gauge for all possible parameter ranges of a model.  
To set the gauge, we make an analogy to 1D chiral symmetric Floquet systems~\cite{asboth2014chiral}, and demand $\nu_{1,2}$ satisfy:
\begin{equation}
\label{eq:chiralwinding0}
\begin{aligned}
&\nu_{1}=n^{0}_{+,0}-n^{0}_{-,0}+n^{0}_{+,1/2}-n^{0}_{-,1/2},
\\
&\nu_{2}=n^{\pi}_{+,0}-n^{\pi}_{-,0}+n^{\pi}_{+,1/2}-n^{\pi}_{-,1/2},
\end{aligned}
\end{equation}
where $n^{0}_{a,w_{x}}$ ($n^{\pi}_{a,w_{x}}$) denotes the number of in-gap eigenstates (e.g., on the left edge) of $\widehat{\mathcal{W}}_{x,2\pi\leftarrow 0}$ ($\widehat{\mathcal{W}}_{x,\pi\leftarrow -\pi}$) at WLO eigenvalue $w_{x},$ and with $m_{x}$-eigenvalue $a$. We call a gauge satisfying Eq.~\eqref{eq:chiralwinding0} a faithful gauge and
will illustrate later in BBH model that we can indeed find such a faithful gauge.

To extract the number of in-gap modes at $0$ and $1/2$ individually, one additional thing to notice is that the number and chirality of in-gap modes of $\widehat{\mathcal{W}}_{x,2\pi\leftarrow 0}$ and $\widehat{\mathcal{W}}_{x,\pi\leftarrow -\pi}$ are not independent. Indeed, we can prove the following statement:

\textbf{Proposition 1:} With mirror symmetry $m_{x}$, if $\widehat{\mathcal{W}}_{x,2\pi\leftarrow 0}$ has an in-gap eigenstate at $0$ ($1/2$) which is also an eigenstate of $m_{x}$, then $\widehat{\mathcal{W}}_{x,\pi\leftarrow -\pi}$ must have an in-gap eigenstate at $0$ ($1/2$) with the same (opposite) $m_{x}$ eigenvalue. 

The proof of Proposition 1 can be found in the Supplementary material \cite{Supp}. According to Proposition 1, $n_{a,w}^{0,\pi}$ satisfy  $n_{a,0}^{0}\PRLequal n_{a,0}^{\pi}$ and $n_{\pm,1/2}^{0}\PRLequal n_{\mp,1/2}^{\pi}$. Substituting these equations into Eq.~\eqref{eq:chiralwinding0}, we can define two topological invariants
\begin{equation}
\label{eq: topoinv}
\begin{aligned}
\nu_{0}&=\frac{\nu_{1}+\nu_{2}}{2}=n^{0}_{+,0}-n^{0}_{-,0}, 
\\
\nu_{1/2}&=\frac{\nu_{1}-\nu_{2}}{2}=n^{0}_{+,1/2}-n^{0}_{-,1/2},
\end{aligned}
\end{equation}
which capture the number of robust in-gap states at $0$ and $1/2$ on a single end, and hence, form a general HOBBC between the boundary Wannier configuration and the bulk topology. 
For example, from above we know the parity of $\nu_{1/2}$ captures the boundary obstruction. 
We also note that a similar discussion also applies for WLOs along $k_{y}$ where there is $m_y$ symmetry.

\begin{figure}[h]
    \centering
    \includegraphics[width=1\columnwidth]{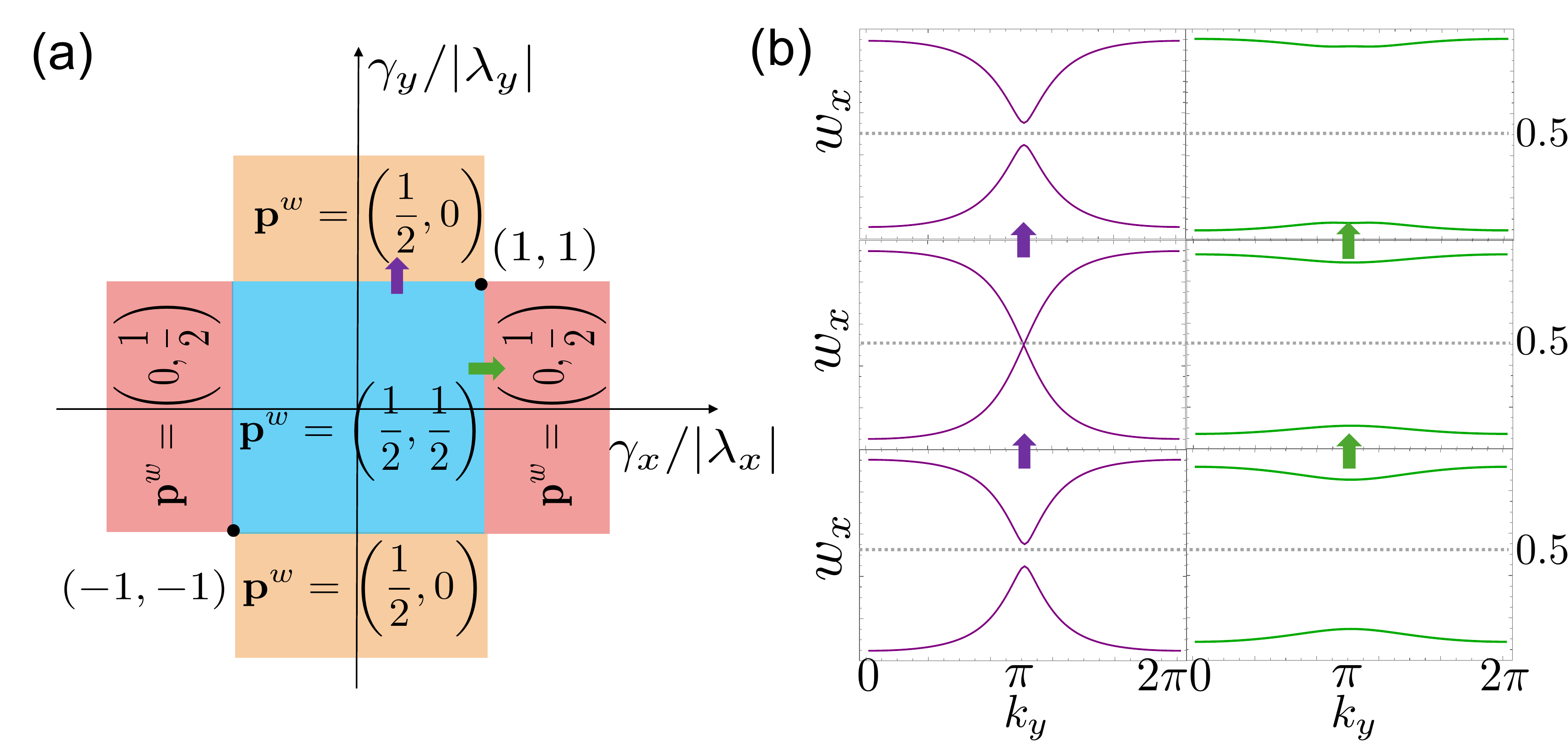}
    \caption{(a) shows the phase diagram of the BBH model, where the color distinguishes topological classes determined by the nested Wilson loops. The white colored region represents the class with  $\mathbf{p}^{w}\PRLequal(0,0)$. The purple and green arrows label transitions where the nested Wilson loops change. The changes of the periodic WLO spectra, $w_{x}$, during the transitions are shown in (b) respectively. Note that in the green subfigure there is no gap closing in the WLO spectrum. }
    \label{fig:BBHphase}
\end{figure}

\emph{Anomalous cases.--} To see how Eq.~\eqref{eq: topoinv} can capture the anomalous cases, i.e., cases where the simple nested Wilson loop analysis fails, we construct an example anomalous case by stacking and coupling multiple copies of BBH models. The Bloch Hamiltonian of the BBH model is \cite{benalcazar2017electric,benalcazar2017quantized}
\begin{equation}
\label{eq: BBH}
\begin{aligned} h^{BBH}(\mathbf{k}) &=\left(\gamma_{x}+\lambda_{x} \cos \left(k_{x}\right)\right) \Gamma_{4}+\lambda_{x} \sin \left(k_{x}\right) \Gamma_{3} \\ &+\left(\gamma_{y}+\lambda_{y} \cos \left(k_{y}\right)\right) \Gamma_{2}+\lambda_{y} \sin \left(k_{y}\right) \Gamma_{1},
\end{aligned}
\end{equation}
where $\Gamma_{i}\PRLequal \PRLminus \tau_{2}\sigma_{i}$ for $i\PRLequal x,y,z$,  $\Gamma_{4}\PRLequal\tau_{1}\sigma_{0}$, and $\tau,\sigma$ are Pauli matrices for the internal degrees of freedom in each unit cell. $h^{BBH}$ has mirror symmetries $m_{i}h^{BBH}(\mathbf{k})m_{i}\PRLequal h^{BBH}(D_{m_{i}}\mathbf{k})$ for $i\PRLequal x,y$, where $m_{x}\PRLequal\tau_{1}\sigma_{3}$ and $m_{y}\PRLequal\tau_{1}\sigma_{1}$, and $D_{m_{i}}$ is the mirror operator in the momentum space. As discussed in Refs. \onlinecite{benalcazar2017quantized,benalcazar2017electric,khalaf2021boundary}, the BBH model has four classes (see Fig.~\ref{fig:BBHphase} (a)) distinguished by the nested Wilson loops $\mathbf{p}^{w}\PRLequal(p^{w}_{x},p^{w}_{y})$, where $p^{w}_{x}$ ($p^{w}_{y}$) is the Zak-Berry phase (divided by $2\pi$) of the Wannier bands of $\widehat{\mathcal{W}}_{y,\pi\leftarrow -\pi}(k_{x})$ ($\widehat{\mathcal{W}}_{x,\pi\leftarrow -\pi}(k_{y})$) living in the spectral region $(0,1/2)$ \footnote{Note that the Zak-Berry phase of the eigenbands in $(0,1/2)$ are actually the same for WLOs with different starting points for the BBH model.}. The class of this model with $\mathbf{p}^{w}\PRLequal(1/2,1/2)$ (blue region in Fig.~\ref{fig:BBHphase} (a)) is the only one having a boundary obstruction. 

\begin{figure}[h]
    \centering
    \includegraphics[width=1\columnwidth]{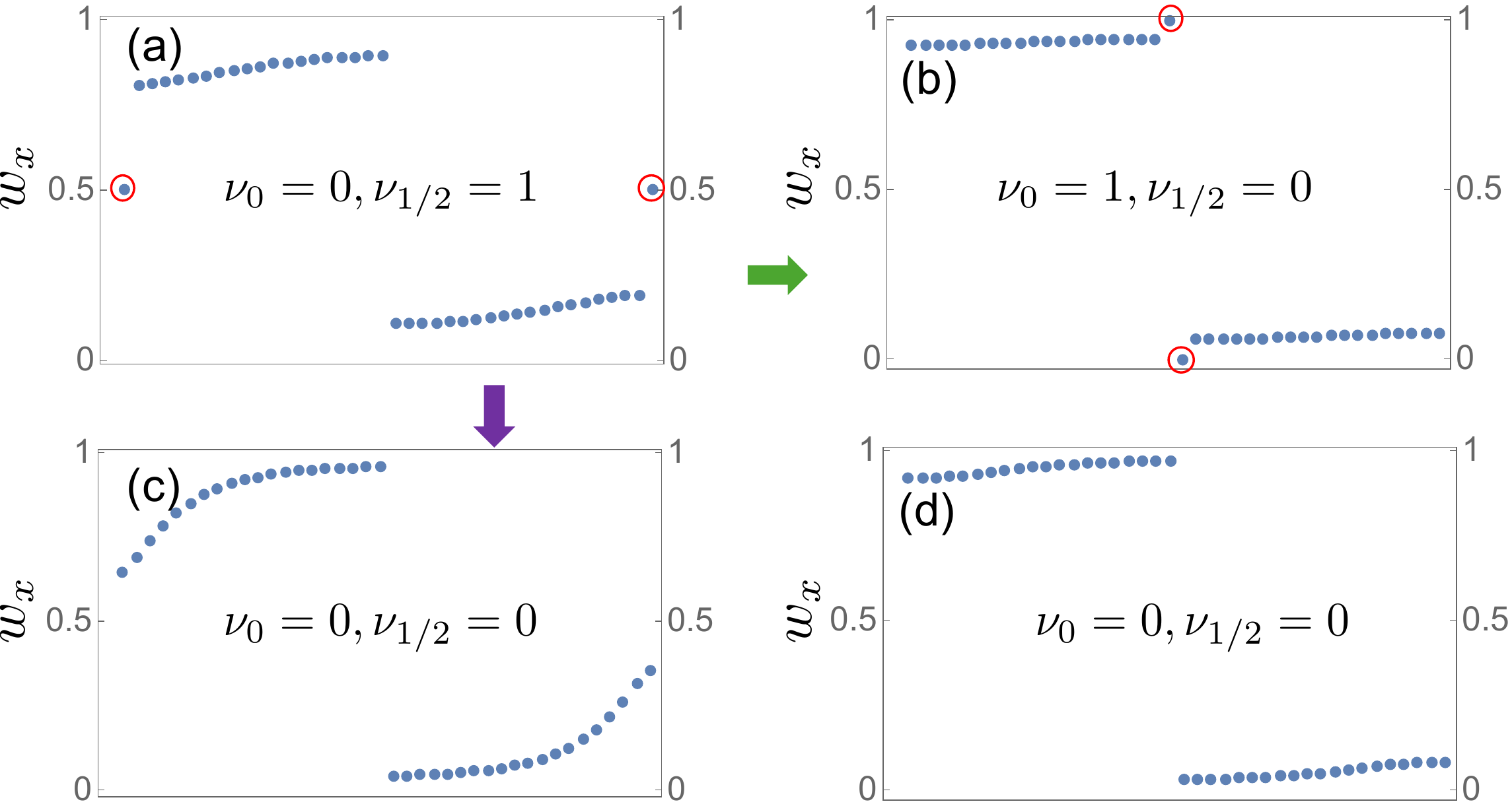}
    \caption{Spectrum of WLOs along $k_{x}$ direction and the corresponding $\nu_{0}$ and $\nu_{1/2}$ for BBH model with parameters (a) $(\gamma_{x}/\lambda_{x},\gamma_{y}/\lambda_{y})=(0.25,0.25)$,
    (b) $(\gamma_{x}/\lambda_{x},\gamma_{y}/\lambda_{y})=(1.25,0.25)$, (c) $(\gamma_{x}/\lambda_{x},\gamma_{y}/\lambda_{y})=(0.25,1.25)$, and (d) $(\gamma_{x}/\lambda_{x},\gamma_{y}/\lambda_{y})=(1.25,1.25)$. The in-gap states highlighted by red circles correspond to Wannier orbitals localized at boundary. The purple and green arrows indicate the same phase transitions as in Fig.~\ref{fig:BBHphase}.
    }
    \label{fig:wspec}
\end{figure}

Now, let us consider stacking (direct summing) two BBH Hamiltonians, one with $\mathbf{p}^{w}\PRLequal(1/2,1/2)$ and another with $\mathbf{p}^{w}\PRLequal(0,1/2).$ The nested Wilson loops are $\mathbb{Z}_2$ quantities so we expect the composite Hamiltonian to have $\mathbf{p}^{w}\PRLequal(1/2,0)$, which implies a trivial phase with no in-gap states in the open boundary WLO spectrum, and thus no boundary obstruction.
However, the open boundary WLO spectrum of the composite Hamiltonian is just a direct combination of (a) and (b) in Fig.~\ref{fig:wspec}. Thus, there is one in-gap state at $0$ and one in-gap state at $1/2$, which is not captured by the nested Wilson loop $\mathbf{p}^{w}\PRLequal(1/2,0).$ Hence combining Hamiltonians does not result in a clear $\mathbb{Z}_2$ group structure for the nested Wilson loops.

With this anomalous case in mind, 
we will now show that $\nu_{0}$ and $\nu_{1/2}$ [c.f. Eq.~\eqref{eq: topoinv}] capture the in-gap boundary modes of the WLO in a cylinder geometry, and thus capture the boundary Wannier configuration.  To see this, let us first discuss the gauge choice for the ZEBs of the WLO. We first note that for the BBH model, the Berry phase of the Wannier bands in $v$ (i.e., the nested Wilson loop) in units of $\pi$ captures the parity of the total number of boundary states sitting in both gaps of the WLO~\cite{benalcazar2017quantized,khalaf2021boundary}. Thus, (i) in order to account for the boundary states in both gaps, the parity of the chiral winding number $\nu_1$ (hence also $\nu_2$ since they must have the same parity for $\nu_{0}$ and $\nu_{1/2}$ to be integers) should be the same as the nested Wilson loop in units of $\pi$. 
Meanwhile, it has been shown that (ii) the parity of chiral winding number $\nu_1$ or $\nu_2$ [cf. Eq.~\eqref{eq: chiralwinding}] is the same as the parity of the Berry phase (in units of $\pi$) of \emph{all} the bands in $v$ (including both Wannier bands and ZEBs)~\cite{mon2014topological}. With (i) and (ii), we see that it is reasonable to choose a gauge such that the ZEBs in $v$ have trivial Berry phase (except possibly at a transition point which we discuss below). 
We provide more details in the Supplementary material \cite{Supp}, but we note that this gauge matches the gauge arrived at via the method developed in Ref.~\onlinecite{Soluyanov2012smooth}.

With this gauge choice, we now calculate and show $\nu_{0}$ and $\nu_{1/2}$ for all four classes of a single BBH model in Fig.~\ref{fig:wspec}, and they indeed capture the in-gap states for all four classes. Thus, this gauge is indeed a faithful gauge. Additionally, we can also observe that $\nu_{0}$ and $\nu_{1/2}$ both have a meaningful $\mathbb{Z}$ group structure, i.e., when enlarging the Hilbert space by a direct sum of several BBH Hamiltonians, we simply add up the $\nu_{0}$ and $\nu_{1/2}$ of all BBH Hamiltonians, and they still correctly predict the bulk-boundary correspondence. In our anomalous example the doubled BBH model yields $\nu_{0}\PRLequal 1$ and $\nu_{1/2}\PRLequal 1$. This fully captures the boundary Wannier orbital configuration, and thus correctly represents the boundary obstruction that is not captured by the nested Wilson loop. Note that the anomalous case discussed here is analogous to the anomalous topological Floquet system shown in Fig.~\ref{fig:floquetanomaly}. 

\emph{Correlations between orthogonal WLOs.--} We have identified that $\nu_{0}$ and $\nu_{1/2}$ correctly classify the anomalous cases beyond the nested Wilson loop, and we will now show they can help understand the correlation between WLOs along two orthogonal directions. As shown in Figs.~\ref{fig:BBHphase} and \ref{fig:wspec}, there are jumps of $\nu_{0}$ and/or $\nu_{1/2}$ at the transitions in the BBH phase diagram indicated by purple and green arrows. The phase transition indicated by the purple arrow happens through a conventional gap closing and reopening in the WLO spectrum (see the left panel of Fig.~\ref{fig:BBHphase}(b)). However, in the transition indicated by the green arrow $\nu_{0}$ and/or $\nu_{1/2}$ have jumps, but there is no gap closing and reopening (see the right panel of Fig.~\ref{fig:BBHphase}(b)) in the WLO spectrum. This is unusual, and we will see that the mechanism of the latter  transition is related to a gap closing in the WLO spectrum along another direction.

\begin{figure}[t!]
    \centering
    \includegraphics[width=1\columnwidth]{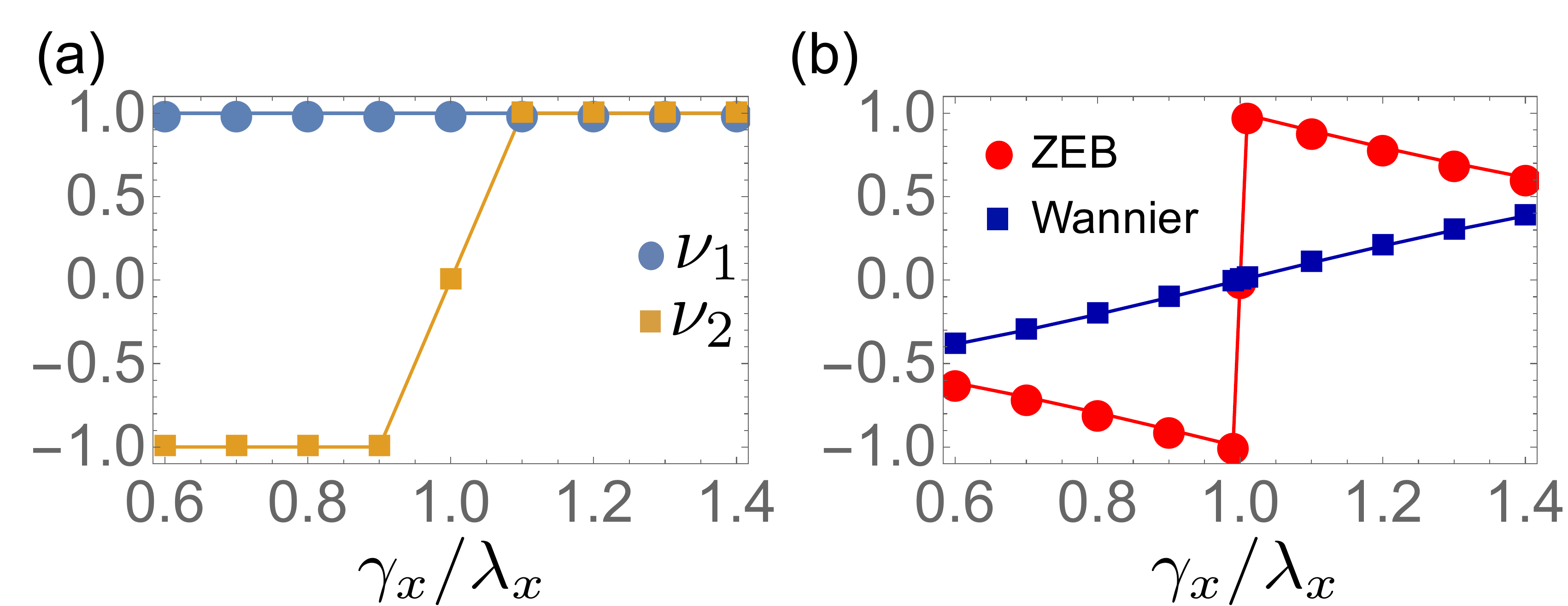}
    \caption{(a) $\nu_{1}$ and $\nu_{2}$ versus $\gamma_{x}/\lambda_{x}$ at $\gamma_{y}/\lambda_{y}=0.25$. (b) The contribution to the chiral winding number $\nu_2$ from  the Wannier band (blue) and ZEB (red) in $v$.  
    }
    \label{fig:phasetrans}
\end{figure}

If we focus on the jump of $\nu_{0}$ and $\nu_{1/2}$ around the green arrow transition point (i.e., $\gamma_{x}/\lambda_{x}\PRLequal 1$), we find that $\nu_{2}$ jumps from $-1$ to $0$ and then to $+1$, while $\nu_{1}$ does not change (see Fig.~\ref{fig:phasetrans}(a)). For 1D chiral symmetric insulators the chiral winding number (mod 2) of bands in $v$ [cf. Eq.~\eqref{eq: chiralwinding}] equals their Zak-Berry phase divided by $\pi$ \cite{mon2014topological}. Hence,
the jump of $\nu_{2}$ from $-1$ to $0$ and then to $+1$ indicates there is a $\pi$ to $0$ and then to $\pi$ transition of the Zak-Berry phase. These Zak-Berry phase transitions occur because the two ZEBs of $\widehat{\mathcal{W}}_{x,\pi\leftarrow -\pi} (k_{y})$ (in the gauge we chose) are smoothly separated into two bands both with trivial Zak-Berry phase when $\gamma_{x}/\lambda_{x}\PRLgreater1$ and $\gamma_{x}/\lambda_{x}\PRLless1$, but both have Zak-Berry phase $\pi$ at $\gamma_{x}/\lambda_{x}\PRLequal 1$. The former is clear from the gauge choice discussion above. To see the latter, we first note that the ZEBs of a WLO always correspond to the unoccupied eigenstates of the Bloch Hamiltonian at the starting momentum of the WLO, since these states are annihilated by the rightmost projector of the WLO. 
In our case, the two ZEBs of $\widehat{\mathcal{W}}_{x,\pi\leftarrow -\pi} (k_{y})$ ($\widehat{\mathcal{W}}_{x,2\pi\leftarrow 0} (k_{y})$) are the two unoccupied bands of $h^{BBH}(k_{x}\PRLequal\pi,k_{y})$ ($h^{BBH}(k_{x}\PRLequal 0,k_{y})$). 
Using this, we see that at the transition point when $\lambda_{x}\PRLequal\gamma_{x}$, $h^{BBH}(k_{x}\PRLequal\pi, k_{y})$ has an  emergent mirror symmetry $\mathcal{M}_y\PRLequal\Gamma_{2}$ in addition to $m_y$. The $\mathcal{M}_y$ eigenvalues are $(++)$ at $k_{y}\PRLequal0$ and $(--)$ at $k_{y}\PRLequal\pi$, which indicates that no matter how we separate the two ZEBs, each of them must have Zak-Berry phase $\pi$ \cite{hughes2011inversion,turner2012quantized,alexandradinata2014wilson}. Hence, at $\lambda_x\PRLequal\gamma_x$ there is a jump of the Zak-Berry phase of the bands in the subspace $v$ from $\pi$ to $0 ({\rm{mod}}  \,2\pi)$ since $v$ contains one of the two ZEBs. This gives us a clue to identify the origin of the jump in $\nu_{2}$ shown in Fig.~\ref{fig:phasetrans}(a). Indeed, in Fig.~\ref{fig:phasetrans}(b) we show the individual contributions to $\nu_{2}$ of the ZEBs and Wannier bands in $v$. We see that the jump is indeed induced by the ZEBs, as we expected from the symmetry-indicated jump in the Zak-Berry phase, and thus a transition of $\nu_i$ can be induced without closing the spectral gap in the WLO. 

This analysis immediately has implications for the WLO constructed from the two \emph{unoccupied} bands of $h^{BBH}(k_{x},k_{y})$ along the orthogonal $k_{y}$-direction. Indeed, the two Wannier band eigenvalues of $\widehat{\mathcal{W}}_{y,\pi\leftarrow -\pi} (k_{x}=\pi)$ correspond to the $\pi$ Zak-Berry phases of the ZEBs for the WLO in the $k_x$-direction, and  are hence degenerate and close the Wannier gap. Since the BBH Hamiltonian has chiral symmetry, this implies that the $\widehat{\mathcal{W}}_{y,\pi\leftarrow -\pi} (k_{x})$ for the occupied bands will become gapless at the transition as well, which matches the phenomenology of orthogonal WLOs found in Ref. \onlinecite{benalcazar2017electric}.
Thus, our results reveal a clear correlation between a WLO in one direction built from occupied bands, and a WLO in the orthogonal direction built from unoccupied bands instead; a distinction which will be more important in Hamiltonians without chiral symmetry.

\begin{figure}[t!]
    \centering
   \includegraphics[width=1\columnwidth]{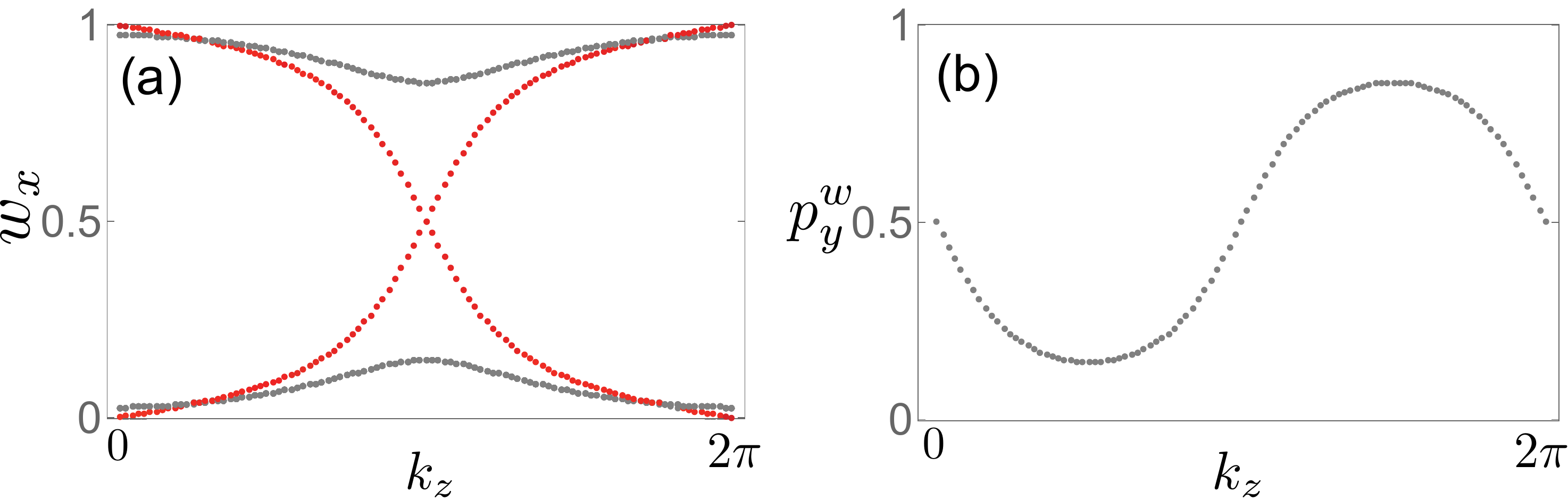}
    \caption{(a) The spectrum of the WLO along the $k_{x}$-direction where the $x$, $z$-directions are periodic, but the $y$-direction is open. The chiral edge states (of two edges) in gaps at $0$ and $1/2$ have been emphasized in red. The winding of chiral edge states on one edge in the spectrum indicates there is a surface Chern number on the $y-z$ surface. (b) The $p^{w}_{y}(t)$ of the 2D Wannier band labeling by $(k_{y},k_{z})$. $\Delta p^{w}_{y}=0$ when $t$ goes from $0$ to $1$ indicates the Chern number of the Wannier band is zero. }
   \label{fig:pump}
\end{figure}

\emph{Generalization to higher dimension.--}Up to now, we have established the correspondence between the WLO multi-gap topology and boundary obstructions in 2D. Remarkably, the WLO multi-gap topology-boundary correspondence can be generalized to higher dimensions. For example, in a 3D model, the surface Chern number, which is manifested as a nontrivial spectral flow of the edge states of an open boundary WLO, is also related to the WLO multi-gap topology. To illustrate, we can construct a 3D model by analogy to a periodic pump of the BBH model from ($\nu_{0}\PRLequal 1,\nu_{1/2}\PRLequal 0$) to ($\nu_{0}\PRLequal 0,\nu_{1/2}\PRLequal 1$) and then back to ($\nu_{0}\PRLequal 1,\nu_{1/2}\PRLequal 0$) as a function of $k_z$:
\begin{equation}
\label{eq: pumpH}
\begin{aligned} & h^{\text{3D}}(\mathbf{k},t) =h^{BBH}(\mathbf{k})+\cos k_{z} \Gamma_{4}
+\sin k_{z} \Gamma_{0},
\end{aligned}
\end{equation}
where $\Gamma_{0}=\tau_{3}\sigma_{0}$. 
As shown in Fig.~\ref{fig:pump}, this model has chiral edge states in both gaps at $0$ and $1/2$ in the spectrum of WLO that gives the nontrivial spectral flow. However, the Chern number of the band between two gaps is zero. This is analogous to the Floquet anomalous Chern insulator \cite{rudner2013anomalous}, which suggests multi-gap topology is needed to capture the in-gap chiral edge states, and thus the Chern number of the 2D surface.  A proper definition of the multi-gap topological invariants for this kind of 3D model is left to future research.

\emph{Conclusions and discussions--} We defined two sets of multi-gap topological invariants for WLOs of 2D mirror symmetric insulators, which fully capture the boundary Wannier orbital configuration, and thus generally represent the higher order bulk-boundary correspondence. Our formalism is useful for understanding anomalous phases and the correlation between WLOs along two orthogonal directions, which are not explained by the  nested Wilson loop. Though illustrated for the BBH model where we have anticommuting $m_{x}$ and $m_{y}$ symmetries, our formalism should widely apply in all mirror symmetric insulators  regardless of the commutation relation of $m_{x}$ and $m_{y}$, as long as there are gapped WLOs in both the occupied and unoccupied space \footnote{It has been shown in Ref.~\cite{benalcazar2017electric} that in a four band model, we must have anticommuting $m_{x}$ and $m_{y}$ to have gapped WLOs. However, in models with bands more than four, gapped WLOs can also appear when we have commuting $m_{x}$ and $m_{y}$. }. Indeed, we show another example in the Supplemental Material.

Our work suggests that there is a close correspondence between the topology of Floquet systems and the WLOs. This implies that the (anomalous) topological phases originally studied in a Floquet context can play a role in static systems through the topology of the WLOs. This understanding of the topology of WLOs brings new insights, and motivates future research on topological phases that can reveal the richness of the
multi-gap topology of WLOs. 
Finally, we point out that our topological invariants can help guide the design and study of 2D metamaterials having boundary obstructions in phononic, photonic, and electric circuit systems where the BBH model has been realized \cite{serra2018observation,peterson2018quantized,imhof2018topolectrical}.

\emph{Acknowledgments.--} X.-Q.~S. acknowledges support from the Gordon and Betty Moore Foundations EPiQS Initiative through Grant GBMF8691. P.~Z. and T.~L.~H. thank the US Office of Naval Research (ONR) Multidisciplinary University Research Initiative (MURI) grant N00014-20- 1-2325 on Robust Photonic Materials with High-Order Topological Protection
for support.

\bibliography{bib}{}
\bibliographystyle{apsrev4-1}



\onecolumngrid
\begin{widetext}
\section{Supplemental Material for: Multi-gap topology of the Wilson loop operator in mirror symmetric insulators}

\beginsupplement

\section{Sec SI. Proof of Proposition 1}
In order to prove Proposition 1, we begin to discuss some important facts about the Wilson line operator when there is a mirror symmetry $m_{x}$.

The usual Wilson line is defined as 
\begin{equation}
\label{eq: Wilsonline1}
\begin{aligned}
&\mathcal{W}_{\mathbf{k}_{2}\leftarrow\mathbf{k}_{1}}  \equiv \mathcal{P} \exp \left[-\int_{\mathbf{k}_{1}}^{\mathbf{k}_{2}} d \boldsymbol{k} \cdot \boldsymbol{A}(\boldsymbol{k})\right]=\lim_{n\rightarrow\infty}F_{\mathbf{k}_{2}=\mathbf{k}_{1}+n\Delta\mathbf{k}}\cdots F_{\mathbf{k}_{1}+\Delta\mathbf{k}}F_{\mathbf{k}_{1}},
\\
& F_{\mathbf{k}}^{ij}=\inp{u^{i}_{\mathbf{k}+\Delta\mathbf{k}}}{u^{j}_{\mathbf{k}}},\quad \boldsymbol{A}_{i j}(\boldsymbol{k})=\left\langle u_{i,\mathbf{k}} \mid \nabla_{k} u_{j, \mathbf{k}}\right\rangle, \quad i, j =1, \ldots, N_{occ},
\end{aligned}
\end{equation}
which is related to the Wilson line operator defined in Eq.(1) in the main text through
\begin{equation}
\mathcal{W}^{ij}_{\mathbf{k}_{2}\leftarrow\mathbf{k}_{1}}=\bra{u_{i}(\mathbf{k}_{2})}\widehat{\mathcal{W}}_{\mathbf{k}_{2}\leftarrow\mathbf{k}_{1}}\ket{u_{j}(\mathbf{k}_{1})}.
\end{equation}
Thus, we can rewrite the Wilson line operator as
\begin{equation}
\label{eq: Wilsonline2}
\widehat{\mathcal{W}}_{\mathbf{k}_{2}\leftarrow\mathbf{k}_{1}}=\mathcal{W}^{ij}_{\mathbf{k}_{2}\leftarrow\mathbf{k}_{1}}\ket{u_{i}(\mathbf{k}_{2})}\bra{u_{j}(\mathbf{k}_{1})}.
\end{equation}
Given $\mathcal{W}_{\mathbf{k}_{2}\leftarrow\mathbf{k}_{1}}$ is unitary in the thermodynamic limit ($n\rightarrow\infty$ in Eq.\eqref{eq: Wilsonline1}), we have
\begin{equation}
\label{eq: qunitarity}
\widehat{\mathcal{W}}_{\mathbf{k}_{2}\leftarrow\mathbf{k}_{1}}\widehat{\mathcal{W}}^{\dag}_{\mathbf{k}_{2}\leftarrow\mathbf{k}_{1}}=P(\mathbf{k}_{2}), \quad \widehat{\mathcal{W}}^{\dag}_{\mathbf{k}_{2}\leftarrow\mathbf{k}_{1}}\widehat{\mathcal{W}}_{\mathbf{k}_{2}\leftarrow\mathbf{k}_{1}}=P(\mathbf{k}_{1}).
\end{equation}

With a mirror symmetry $m_{x}$, there will be a constraint on the Wilson line operator:
\begin{equation}
\label{eq: mirrorWLO1}
m_{x}\widehat{\mathcal{W}}_{x,0\leftarrow -\pi}m_{x}=\widehat{\mathcal{W}}_{x,0\leftarrow \pi}\equiv \widehat{\mathcal{W}}^{\dag}_{x,\pi\leftarrow 0 },
\end{equation}
from which we can write the WLOs $\widehat{\mathcal{W}}^{\dag}_{x,2\pi\leftarrow 0}$ and $\widehat{\mathcal{W}}^{\dag}_{x,\pi\leftarrow -\pi}$ as
\begin{equation}
\label{eq: WLO1}
\begin{aligned}
&\widehat{\mathcal{W}}^{\dag}_{x,2\pi\leftarrow 0} =m_{x}\widehat{\mathcal{W}}^{\dag}_{x,\pi\leftarrow 0} m_{x}\widehat{\mathcal{W}}_{x,\pi\leftarrow 0}, 
\\
&\widehat{\mathcal{W}}^{\dag}_{x,\pi\leftarrow -\pi} =\widehat{\mathcal{W}}_{x,\pi\leftarrow 0}m_{x}\widehat{\mathcal{W}}^{\dag}_{x,\pi\leftarrow 0} m_{x}.
\end{aligned}
\end{equation}
From Eq.\eqref{eq: WLO1}, it is straightforward that $\widehat{\mathcal{W}}_{x,\pi\leftarrow -\pi}\widehat{\mathcal{W}}_{x,\pi\leftarrow 0}=\widehat{\mathcal{W}}_{x,\pi\leftarrow 0}\widehat{\mathcal{W}}_{x,2\pi\leftarrow 0}$. Then if $\widehat{\mathcal{W}}^{\dag}_{x,2\pi\leftarrow 0}$ has an in-gap eigenstate $\ket{\psi}$ that satisfies $\widehat{\mathcal{W}}^{\dag}_{x,2\pi\leftarrow 0}\ket{\psi}=e^{-i2\pi w_{x}}$ with $w_{x}=0,1/2$, and $m_{x}\ket{\psi}=e^{i\gamma}\ket{\psi}$ with $\gamma=0,\pi$, we have
\begin{equation}
\label{eq: result1}
\widehat{\mathcal{W}}_{x,\pi\leftarrow -\pi}\widehat{\mathcal{W}}_{x,\pi\leftarrow 0}\ket{\psi} =\widehat{\mathcal{W}}_{x,\pi\leftarrow 0}\widehat{\mathcal{W}}_{x,2\pi\leftarrow 0}\ket{\psi}= e^{-i 2\pi w_{x}}\widehat{\mathcal{W}}_{x,\pi\leftarrow 0}\ket{\psi},
\end{equation}
and
\begin{equation}
\label{eq: result2}
\begin{aligned}
&m_{x}\widehat{\mathcal{W}}_{x,\pi\leftarrow 0}\ket{\psi}=(P(\pi,k_{y})+Q(\pi,k_{y}))m_{x}\widehat{\mathcal{W}}_{x,\pi\leftarrow 0}\ket{\psi}
\\
&=(\widehat{\mathcal{W}}_{x,\pi\leftarrow 0}m_{x}m_{x}\widehat{\mathcal{W}}^{\dag}_{x,\pi\leftarrow 0})m_{x}\widehat{\mathcal{W}}_{x,\pi\leftarrow 0}\ket{\psi}=e^{i(\gamma-2\pi w_{x})}\widehat{\mathcal{W}}_{x,\pi\leftarrow 0}\ket{\psi},
\end{aligned}
\end{equation}
where the second step of the second equation used $m_{x}^{2}=1$, $[m_{x},Q(\pi,k_{y})]=0$, $P(\pi,k_{y})+Q(\pi,k_{y})=\mathbbm{1}$, $Q(\pi,k_{y})P(\pi,k_{y})=0$ (thus $Q(\pi,k_{y})\widehat{\mathcal{W}}_{x,\pi\leftarrow 0}=0$ ), and Eq.\eqref{eq: qunitarity}. Eq.\eqref{eq: result1} and Eq.\eqref{eq: result2} together give the important statement:  if $\widehat{\mathcal{W}}_{x,2\pi\leftarrow 0}$ has an in-gap eigenstate at $0$ ($1/2$) which is also an eigenstate of $m_{x}$, then $\widehat{\mathcal{W}}_{x,\pi\leftarrow -\pi}$ must have an in-gap eigenstate at $0$ ($1/2$) with the same (opposite) $m_{x}$ eigenvalue.

\section{Sec SII. Proof of Eq.(4)}
Without loss of generality, let us focus on a WLO along $k_x$ in a system with mirror symmetry $m_{x}$ along $x$ direction, of which there are even number of zero eigenvalue bands (ZEBs). To separate the ZEBs mirror symmetrically, we first diagonalize $Pm_{x}P$ where $P$ is the projector on the subspace spanned by all ZEBs, and find a smooth gauge to obtain two sets of bands, $\{\ket{\psi^{+}_{j}(k_{y})}\}$ and $\{\ket{\psi^{-}_{j}(k_{y})}\}$, which have $m_{x}$ eigenvalues of $+1$ and $-1$, respectively. Note that these two sets of states live in two orthogonal subspace (with the same dimension) labelled by $m_{x}$ eigenvalue in the sense that
\begin{equation}
\label{eq: c}
\inp{\psi_{j}^{+}(k_{y1})}{\psi_{l}^{-}(k_{y2})}=0
\end{equation}
for any $j,l$ and $k_{y1}, k_{y2}$. Mathematically, given $m_{x}$ is an unitary operator such that $m_{x}m_{x}^{\dag}=m_{x}^{\dag}m_{x}=\mathbbm{1}$, Eq.\eqref{eq: c} can be proved through
\begin{equation}
\inp{\psi_{j}^{+}(k_{y1})}{\psi_{l}^{-}(k_{y2})}=\bra{\psi_{j}^{+}(k_{y1})}m_{x}^{\dag}m_{x}\ket{\psi_{l}^{-}(k_{y2})}= \inp{m_{x}\psi_{j}^{+}(k_{y1})}{m_{x}\psi_{l}^{-}(k_{y2})}=-\inp{\psi_{j}^{+}(k_{y1})}{\psi_{l}^{-}(k_{y2})}.
\end{equation}

For a given $j$, we choose a $l$ and put one of $(\ket{\psi^{+}_{j}(k_{y})}\pm \ket{\psi^{-}_{l}(k_{y})})/\sqrt{2}$ into $v$. In this way, $v$ and its complement are related by $m_{x}$, and for each choice, we have a mirror symmetric separation. Now we prove Eq.(3) in the main text does not depend on there choices. The contribution to Eq.(3) from the ZEBs is given by a integral over $k_{y}$ of
\begin{equation}
\label{eq: c1}
\begin{aligned}
 &\frac{1}{2}\sum_{\{j,l\}} i (\bra{\psi^{+}_{j}(k_{y})}\pm \bra{\psi^{-}_{l}(k_{y})})m_{x}\partial_{k_{y}}(\ket{\psi^{+}_{j}(k_{y})}\pm \ket{\psi^{-}_{l}(k_{y})})
 \\
 &=\frac{1}{2}\sum_{\{j,l\}} i (\bra{\psi^{+}_{j}(k_{y})}\mp \bra{\psi^{-}_{l}(k_{y})})\partial_{k_{y}}(\ket{\psi^{+}_{j}(k_{y})}\pm \ket{\psi^{-}_{l}(k_{y})}),
\end{aligned}
\end{equation}
where $\sum_{\{j,l\}}$ means the sum over all states we put in $v$ with a fixed choice of $l$ for a given $j$. For each $\{j,l\}$ pair, we still have the freedom to choose $(\ket{\psi^{+}_{j}(k_{y})}+ \ket{\psi^{-}_{l}(k_{y})})/\sqrt{2}$ or $(\ket{\psi^{+}_{j}(k_{y})}- \ket{\psi^{-}_{l}(k_{y})})/\sqrt{2}$ to be in $v$. If we expand Eq.\eqref{eq: c1}, we have
\begin{equation}
\label{eq: c2}
\frac{i}{2}\sum_{\{j,l\}}\left(\bra{\psi^{+}_{j}}\partial_{k_y}\ket{\psi^{+}_{j}}-\bra{\psi^{-}_{l}}\partial_{k_y}\ket{\psi^{-}_{l}}\mp\bra{\psi^{-}_{l}}\partial_{k_y}\ket{\psi^{+}_{j}}\pm \bra{\psi^{+}_{j}}\partial_{k_y}\ket{\psi^{-}_{l}}\right).
\end{equation}
With Eq.\eqref{eq: c}, it is straightforward to see the last two terms of Eq.\eqref{eq: c2} are zero. Then, whatever ${j,l}$ pairing we choose, Eq.\eqref{eq: c2} is always the same and takes the form
\begin{equation}
\label{eq: c3}
\frac{i}{2}\left(\sum_{j}\bra{\psi^{+}_{j}}\partial_{k_y}\ket{\psi^{+}_{j}}-
\sum_{l}\bra{\psi^{-}_{l}}\partial_{k_y}\ket{\psi^{-}_{l}}\right),
\end{equation}
which is just Eq.(4).

\section{Sec SIII. Gauge choice of the BBH model}
Using the results in Sec. SII, the Zak-Berry phase of one set of ZEBs after separation is
\begin{equation}
\begin{aligned}
&\int dk_{y} \frac{1}{2}\sum_{\{j,l\}} i (\bra{\psi^{+}_{j}(k_{y})}\pm \bra{\psi^{-}_{l}(k_{y})})\partial_{k_{y}}(\ket{\psi^{+}_{j}(k_{y})}\pm \ket{\psi^{-}_{l}(k_{y})})
\\
&=\int dk_{y} \frac{i}{2}\left(\sum_{j}\bra{\psi^{+}_{j}}\partial_{k_y}\ket{\psi^{+}_{j}}+
\sum_{l}\bra{\psi^{-}_{l}}\partial_{k_y}\ket{\psi^{-}_{l}}\right),
\end{aligned}
\end{equation}
where $\frac{1}{2\pi}\int dk_{y} i\sum_{j}\bra{\psi^{\pm}_{j}}\partial_{k_y}\ket{\psi^{\pm}_{j}}$ gives the sum of the Wannier centers along $y$ direction of all $m_{x}=\pm 1$ states at mirror invariant $k_{x}$. With the gauge such that the Zak-Berry phase of each $\ket{\psi^{\pm}_{j}(k_{y})}$ is in $(-\pi,\pi]$,  the $m_{y}$ symmetry enforces that states must come in pair that have opposite Wannier centers along $y$, i.e., $\int dk_{y} i\left(\sum_{j}\bra{\psi^{+}_{j}}\partial_{k_y}\ket{\psi^{+}_{j}}=- \int dk_{y} i
\sum_{l}\bra{\psi^{-}_{l}}\partial_{k_y}\ket{\psi^{-}_{l}}\right) $. Thus,
\begin{equation}
\label{eq:berry}
\begin{aligned}
\int dk_{y} \frac{i}{2}\left(\sum_{j}\bra{\psi^{+}_{j}}\partial_{k_y}\ket{\psi^{+}_{j}}+
\sum_{l}\bra{\psi^{-}_{l}}\partial_{k_y}\ket{\psi^{-}_{l}}\right)=0,
\end{aligned}
\end{equation}
except when there is gap closing at $1/2$ in the spectrum of the WLO constructed from unoccupied subspace along $k_{y}$ direction at mirror invariant $k_{x}$. The proof is still true if we exchange $k_{x}$ and $k_{y}$.



\section{Sec SIV. Multi-gap invariants for the model in Ref. 35}
In Ref.~\onlinecite{Franca:2018}, the authors constructed a model for a 2D array
of Majorana nanowires, of which the Hamiltonian is
\begin{equation}
\label{eq: Majorana}
\begin{aligned}
&H(\mathbf{k})=\left[2 t_{x}\left(1-\cos k_{x}\right)-\mu\right] \tau_{z} \sigma_{0} \eta_{0}+V_{z} \tau_{0} \sigma_{z} \eta_{0}+\Delta \tau_{x} \sigma_{0} \eta_{0}+\alpha \sin k_{x} \tau_{z} \sigma_{y} \eta_{0}- \\
&\beta_{1} \tau_{z} \sigma_{x} \eta_{y}-\beta_{2} \sin k_{y} \tau_{z} \sigma_{x} \eta_{x}+\beta_{2} \cos k_{y} \tau_{z} \sigma_{x} \eta_{y},
\end{aligned}
\end{equation}
where $\tau$, $\sigma$, and $\eta$ are Pauli matrices acting on the particle-hole, spin, and wire space, respectively. With the parameters $\alpha=3.7$, $t_x=1.7$, $\mu=-0.9$, $\Delta=1.6$, $V_z=2.7$, $\beta_{1}=0.8$, and $\beta_{2}=6.2$, Eq. \eqref{eq: Majorana} has anomalous WLOs (constructed from occupied space) along $k_{x}$ direction. 

As discussed in the main text, we use the algorithm in Ref.~\onlinecite{Soluyanov2012smooth} to do gauge smoothing of the ZEBs, which automatically gives the gauge such that all $\ket{\psi_{+}}$ and $\ket{\psi_{-}}$ have a Berry phase in $(-\pi,\pi]$. Numerical calculations of the WLOs (constructed from occupied space) along $k_{x}$ direction give $\nu_{1}=-2$ and $\nu_{2}=0$. This leads to $\nu_{0}=-1$ and $\nu_{1/2}=-1$, which correctly captures the in-gap boundary states (see Fig.\ref{fig:majoranamodel} (a)). The WLOs (constructed from occupied space) along $k_{y}$ direction of Eq. \eqref{eq: Majorana} with the given parameters are normal. Numerical calculations give $\nu_{1}=1$ and $\nu_{2}=-1$, and thus $\nu_{0}=0$ and $\nu_{1/2}=1$, which correctly captures the in-gap boundary states (see Fig.\ref{fig:majoranamodel} (b)).

\begin{figure}[h]
    \centering
    \includegraphics[width=0.7\columnwidth]{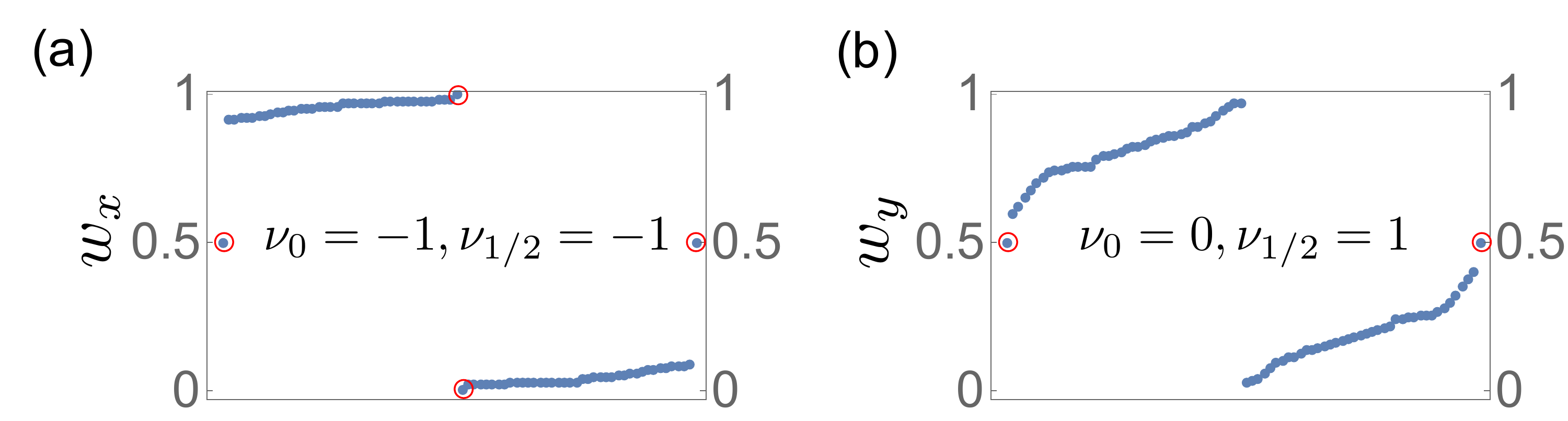}
    \caption{Spectrum of WLOs constructed from occupied bands along (a) $k_{x}$ direction and (b) $k_{y}$ direction, and the corresponding $\nu_{0}$ and $\nu_{1/2}$ for models in Eq.\eqref{eq: Majorana} with parameters $\alpha=3.7$, $t_x=1.7$, $\mu=-0.9$, $\Delta=1.6$, $V_z=2.7$, $\beta_{1}=0.8$, and $\beta_{2}=6.2$. The in-gap states highlighted by red circles correspond to Wannier orbitals localized at boundary.}
    \label{fig:majoranamodel}
\end{figure}

\end{widetext}

\end{document}